\documentclass[aps,showpacs,floatfix,preprintnumbers,nofootinbib,11pt,prd]{revtex4-1}

\usepackage{amsthm,amsmath,amssymb,cancel,hyperref}
\usepackage{xcolor}
\usepackage[textwidth=6.3in,top=1.3in,bottom=1.3in]{geometry}

\usepackage{bm}
\usepackage{latexsym}
\usepackage{amsmath}
\usepackage{graphicx}
\usepackage{caption}
\usepackage{subcaption}

\usepackage{amsmath}
\numberwithin{equation}{section}

\newcommand{\kmsMpc}{\, \text{km}\,\text{s}^{-1}\, \text{Mpc}^{-1}}

\begin{document}
%\thispagestyle{empty}
%\nopagebreak

\title{New Early Dark Energy is compatible with current LSS data}

\author{Florian Niedermann}
\email{niedermann@cp3.sdu.dk}
\author{Martin S.~Sloth} 
\email{sloth@cp3.sdu.dk}
\affiliation{CP$^3$-Origins, Center for Cosmology and Particle Physics Phenomenology \\ University of Southern Denmark, Campusvej 55, 5230 Odense M, Denmark}

\preprint{\tt CP3-Origins-2020-11 DNRF90}

\begin{abstract}
Recently a full-shape analysis of large-scale structure (LSS) data was employed to provide new constraints on a class of Early Dark Energy (EDE) models. In this note, we derive similar constraints on New Early Dark Energy (NEDE) using the publicly available {\tt PyBird} code, which makes use of the effective field theory of LSS. We study the NEDE base model with the fraction of NEDE and the trigger field mass as two additional parameters allowed to vary freely while making simplifying assumptions about the decaying fluid sector.  Including the full-shape analysis of LSS together with measurements of the cosmic microwave background (CMB), baryonic acoustic oscillations (BAO) and supernovae (SN) data, we report $ H_0= 71.2 \pm 1.0 \kmsMpc $ ($68 \%$ C.L.) together with a $\simeq 4 \, \sigma$ evidence for a non-vanishing fraction of NEDE. This is an insignificant change to the value previously found without full-shape LSS data, $ H_0= {71.4 \pm 1.0  \kmsMpc} $ ($68 \%$ C.L.).  As a result, while the NEDE fit cannot be improved upon the inclusion of additional LSS data, it is also not adversely affected by it, making it compatible with current constraints from LSS data. In fact, we find evidence that the effective field theory of LSS acts in favor of NEDE.
\end{abstract}

\maketitle

\section{Introduction}
Recent direct measurements of the expansion rate of the universe using Type Ia SN as standard candles (SH0ES~\cite{Riess:2019cxk}) are in tension with the expansion rate inferred form the CMB~\cite{Aghanim:2018eyx} when assuming the standard $\Lambda$CDM cosmological model (for recent reviews see \cite{Bernal:2016gxb,Knox:2019rjx,Verde:2019ivm,Riess:2020sih}).\footnote{A similarly high value was reported based on time delays caused by strong gravitational lenses (H0LiCOW~\cite{Wong:2019kwg}); however, these measurements rely sensitively on assumptions about the mass density profile of elliptical galaxies~\cite{Birrer:2020tax}.} It is a hot subject of discussion whether unaccounted for systematical effects in astronomical distance measurements are responsible for this discrepancy or whether we have to refine our understanding of the history of the universe by going beyond the $\Lambda$CDM model~\cite{Freedman:2019jwv,Freedman:2020dne}. It turns out that cosmological measurements have reached a precision where modifying the history of the universe to bring new concordance between CMB and direct measurements of the expansion rate is very difficult without introducing new tensions between different data sets. As theorists looking for a new concordance model to replace $\Lambda$CDM, we are truly experiencing that we have entered the era of precision cosmology. If we try to modify the late-time history to account for the higher value of $H_0$ measured today by SH0ES, then we quickly run into tension with BAO measurements \cite{Bernal:2016gxb,Aylor:2018drw,Verde:2016ccp,Knox:2019rjx,Arendse:2019itb}.  On the other hand, an extra component of dark energy, which decays away shortly before recombination has proved more promising as a possible solution~\cite{Poulin:2018dzj,Poulin:2018cxd,Smith:2019ihp,Lin:2019qug,Kaloper:2019lpl,Alexander:2019rsc,Hardy:2019apu,Sakstein:2019fmf,1798362, Gonzalez:2020fdy}. The early dark energy (EDE) proposal suggests that the early dark energy is stored in a slow-rolling scalar field and decays away as the scalar field approaches the bottom of its potential and picks up speed, similar to how inflation ends in slow-roll inflation. However, in order to satisfy phenomenological constraints the potential has to be relatively fine-tuned. But more seriously, recent fits including a full-shape analysis of LSS data have challenged the ability of EDE to solve the Hubble tension at all \cite{Hill:2020osr,Ivanov:2020ril,DAmico:2019fhj}.

Many of the issues with EDE are avoided in the more recent new early dark energy (NEDE) proposal~\cite{Niedermann:2019olb,Niedermann:2020dwg}. Here, the early dark energy decays away in a first order phase transition, and the free energy released is partially converted into small-scale anisotropic stress (behaving similar to a stiff fluid on large scales) and gravitational radiation, which provides a good fit to all cosmological data, including the recent measurements of $H_0$ by SH0ES. It remains however to be seen what a full-shape analysis of LSS data will imply for NEDE to serve as a new concordance model. The purpose of this short note is to provide such an analysis.

The full-shape analysis of the matter power spectrum within EDE was carried out in \cite{Ivanov:2020ril} and~\cite{DAmico:2020ods} using the effective field theory of large-scale structure\footnote{The EFTofLSS was first formulated in \cite{Baumann:2010tm,Carrasco:2012cv,Porto:2013qua} and later used to compute the dark matter power spectrum \cite{Carrasco:2013sva,Carrasco:2013mua,Carroll:2013oxa,Senatore:2014via,Baldauf:2015zga,Foreman:2015lca,Baldauf:2015aha,Cataneo:2016suz,Lewandowski:2017kes,Konstandin:2019bay,Pajer:2013jj,Abolhasani:2015mra,Mercolli:2013bsa}. An IR-resummed version of the EFTofLSS was then able to reproduce the BAO peak \cite{Senatore:2014vja, Lewandowski:2014rca, Baldauf:2015xfa, Blas:2016sfa,Senatore:2017pbn, Lewandowski:2018ywf}. For a more complete account of all related research see for example the references provided in \cite{DAmico:2019fhj}.}
 (EFTofLSS) applied to BOSS/SDSS data~\cite{DAmico:2019fhj,Ivanov:2019pdj,Colas:2019ret}. In particular, in  \cite{DAmico:2020ods} the code {\tt PyBird} was used for the full-shape analysis of LSS data and also made public by the same collaboration. This has enabled us to use the same code to repeat their analysis for NEDE. We therefore employ {\tt PyBird} to analyze the full-shape of the LSS power spectrum, and use it alongside CMB, (small-$z$ and large-$z$) BAO and supernovae data to constrain NEDE. We have tested our implementation of it on $\Lambda$CDM and $w$CDM where we find agreement with \cite{DAmico:2020ods} and~\cite{DAmico:2020kxu}, respectively.

Below we will provide a short review of the NEDE model, following~\cite{Niedermann:2019olb,Niedermann:2020dwg}, and then discuss the data analysis and results.

\section{New Early Dark Energy}

\subsection{Summary of the Model}
NEDE is associated with the false vacuum energy of a two-component scalar field $(\psi, \phi)$ that undergoes a first order phase transition. The corresponding potential reads\footnote{As compared to \cite{Niedermann:2020dwg}, we have set $\beta = 1$ which can always be achieved by rescaling $M$. This generic potential has been studied before in an inflationary context in \cite{Linde:1990gz,Adams:1990ds,Copeland:1994vg}.}
\begin{align}\label{eq:action2}
V(\psi, \phi) = \frac{\lambda}{4} \, \psi^4  + \frac{1}{2} M^2 \psi^2 - \frac{1}{3} \alpha M \psi^3 
+ \frac{1}{2} m^2 \phi^2 +\frac{1}{2} \tilde{\lambda} \, \phi^2 \psi^2 \,,
\end{align}
where the parameters $\lambda$, $\tilde \lambda$ and $\alpha$ are positive and dimensionless, and we assume that $\lambda / \alpha^2 <  1/4 $ for the potential to have a non-trivial vacuum structure. In particular, the true vacuum corresponds to $( \psi, \phi)_\text{True} =  \left(\frac{M}{2 \lambda}\,\left[\alpha + \sqrt{\alpha^2-4 \lambda }\right] ,0 \right)$. With these definitions, the NEDE background energy density at decay time $t_*$ is $\bar{\rho}_\text{NEDE}(t_*) = V\left(\psi_\text{False},\phi(t_*)\right) - V(\psi_\text{True},0)$, corresponding to a fraction of the total energy density $\bar{\rho}$ given by $f_\text{NEDE} = \bar{\rho}_\text{NEDE}(t_*) / \bar{\rho} (t_*)$. The definition of $f_\text{NEDE}$ does not include the kinetic energy of the fields, which is suppressed before the decay. This additional energy component then leads to an increase in the Hubble parameter $H(t)$, prior to recombination, which in turn reduces the comoving sound horizon $r_s(z_\text{rec}) = \int_{z_\text{rec}}^\infty dz \, v(z) / H(z)$, with $v(z)$ denoting the sound speed in the photon-baryon fluid. This alone would shift the angular position of the first peak in the CMB power spectrum, $\theta_\text{rec} = r_s(z_\text{rec})/D_\text{rec}$, which is highly constrained. The change in $r_s(z_\text{rec})$ is, however, compensated by simultaneously lowering the comoving distance to the surface of last scattering $D_\text{rec}$, such that $\theta_\text{rec} $ remains unchanged. As $D_\text{rec} \propto 1/ H_0$, this is achieved by an increase in the Hubble parameter $H_0$, which at the same time resolves the tension between its local and CMB inferred value. For this mechanism to work, it is crucial that the NEDE energy component decays around recombination to avoid over-closing the universe. In our case this decay is triggered by the ultralight scalar field $\phi(t)$. It traces an almost flat direction in field space, corresponding to the mass scale $m \sim 10^{-27} \text{eV}$, whereas $\psi$ has a much heavier mass, $M \sim 0.1 \,\text{eV} \gg m$, setting the scale of NEDE. This huge hierarchy can be stabilized against quantum corrections by imposing~\cite{Niedermann:2020dwg} $\tilde \lambda < \mathcal{O}(1) \times 10^3 \, m^2 / M^2 $. Initially, the field is frozen due to the Hubble friction and prevented from tunneling to the true minimum by a high potential barrier in $\psi$ direction, explicitly $( \psi, \phi)_\text{ini} = (0 , \phi_\text{ini}) \simeq \mathrm{const}$. Once the Hubble drag gets released, $\phi$ starts rolling, thereby decreasing the potential barrier `seen' by $\psi$ and triggering the phase transition. This happens within one Hubble time and before $\phi$ reaches its minimum. The exact timing depends on the details of the potential, but falls in the range\footnote{As argued in \cite{Niedermann:2020dwg}, the upper bound tightens to $\lesssim 0.21$ when suppressing oscillations of $\phi$ around the true vacuum. } $0.18 < H(t_*) / m  < \mathcal{O}(1) $, where the decay time $t_*$ is implicitly determined trough the `trigger parameter' $H(t_*) / m$. The lower bound corresponds to the point of maximal tunneling probability (when $\phi$ crosses zero for the first time) and the upper bound ensures that the Hubble drag has been released. For $\alpha = \mathcal{O}(1)$ the tunneling rate is $\Gamma(t) \sim  M^4 e^{-S_E(t)}$, where $S_E(t)$ is the Euclidian action evaluated at the `bounce solution'. Its time dependence is inherited from the trigger field $\phi(t)$ that scans the potential. The inverse duration of the phase transition $\bar{\beta} \simeq \dot \Gamma /\Gamma$ (approximating $\Gamma \propto e^{\bar{\beta} t }$) was then found to be~\cite{Niedermann:2020dwg}\footnote{This expression also controls the amplitude of the gravitational wave signal produced during the phase transition. In~\cite{Niedermann:2020dwg}, it is argued to be marginally compatible with the peak sensitivity of future pulsar timing arrays in the limit where $H \bar{\beta}^{-1} \to 1$.}
\begin{align}
\label{eq:beta_final}
H(t_*) \bar{\beta}^{-1} & = \mathcal{O}(1) \times \, 10^{-3}\, \left(\frac{f_\text{NEDE}}{0.1}\right)^{1/2}\, \left(11 \alpha^2 /9 -1 \right)^{-1/2} \left(\frac{M_{pl}/\phi_\text{ini}}{10^4} \right)\, \left(\frac{H(t_*)/m}{0.2} \right)\, \left( \frac{\lambda}{0.01}\right)^{3/2}\,,
\end{align}
provided $\tilde \lambda$ saturates its naturalness bounds and the quartic coupling is sufficiently weak, $\lambda < 0.02$. Here,  we used that $S_E(t_*) \simeq 250 $ which follows from the percolation condition $\Gamma (t_*) / H^4 \gtrsim 1 $. It is fulfilled when bubble nucleation becomes efficient, i.e.~there is more than one nucleation event per Hubble time and volume.  For the above suggested parameter choices, we therefore find that the phase transition happens on a time scale that is short compared to the Hubble expansion.  After the transition, the space is filled with a condensate of colliding bubble walls. This state is dominated by small-scale anisotropic stress and expected to behave on large scales like a fluid dominated by kinetic rather than potential energy, and, therefore, it decays quicker than radiation.  As a result, our microscopic model can be described in terms of an effective cosmological fluid, dubbed NEDE fluid, which first (before $t_*$) behaves like vacuum energy but then (after $t_*$) redshifts away with an equation of state parameter $w_\text{NEDE}(t) > 1/3$. In other words, our effective model underlies the assumption that the effect of small-scale non-linearities can be captured in terms of a cosmic fluid. Ultimately, the fluid parameters such as $w_\text{NEDE}$ are determined by the microscopic details of our underlying field theory model. A priori, $w_\text{NEDE}$ is time-dependent, but we will approximate it as a constant. This is justified because NEDE can impact cosmological observables only in a short redshift window around its decay time; explicitly, we take
\begin{align}
w_\text{NEDE}(t) = 
	\begin{cases}
		-1   &\text{for} \quad t< t_* \,, \\ 
		w_\text{NEDE}(t_*)  &\text{for}\quad t \geq t_* \,.
	\end{cases}
\end{align} 
At the background level, NEDE is therefore described in terms of four parameters, the fraction of NEDE at decay time $f_\text{NEDE}$, the mass of the trigger field, $m$, the trigger parameter $H(t_*)/m$ (which together with $m$ fixes the decay redshift $z_*$), and the equation of state of the NEDE fluid right after percolation has been completed $w_\text{NEDE}(t^*)$. The energy density related to the trigger field is always sub-dominant. As a result, cosmological observables are not sensitive to the initial value of $\phi$ as long as it is sub-Planckian.
Perturbations in the NEDE fluid are generated after the phase transition. They arise from adiabatic perturbations of the trigger field $\delta \phi(t,\mathbf{x}) $, which cause spatial variations of the decay time. The NEDE density contrast and velocity divergence as a function of $k$ right after the decay are~\cite{Niedermann:2020dwg}
\begin{subequations}
\label{eq:EDE_ini}
\begin{align}
\delta_\text{NEDE} (t_*, k)  &= - 3 \left[ 1 + w_\text{NEDE}(t_*)\right] H(t_*) \frac{\delta \phi(t_*,k)}{\dot{\phi}(t_*)} \; ,\\
\theta_\text{NEDE}(t_*,k) &= \frac{k^2}{a(t_*)} \frac{\delta \phi(t_*,k)}{\dot{\bar{\phi}}(t_*)} \,.
\end{align}
\end{subequations}
These initial values are then propagated forward in time using the adiabatic perturbation equations of a generic fluid~\cite{Ma:1995ey}. We note that for adiabatic perturbations, the matching equations are independent of the initial value of $\phi$.\footnote{A more detailed discussion of this point is provided within the methodology part of~\cite{Niedermann:2020dwg}.} This system can be generalized by allowing for a non-vanishing viscosity parameter and a rest-frame sound speed that deviates from the adiabatic sound speed. These extensions are further investigated in~\cite{Niedermann:2020dwg}. Here, we limit the discussion to the simplest case with $f_\text{NEDE}$ and $m$ as two additional parameters while setting explicitly $w_\text{NEDE}=2/3$.

\subsection{{Linear Matter Power Spectrum}}\label{sec:sigma8}

\begin{figure}
     \centering
          \begin{subfigure}[b]{0.49\textwidth}
         \centering
         \includegraphics[width=\textwidth]{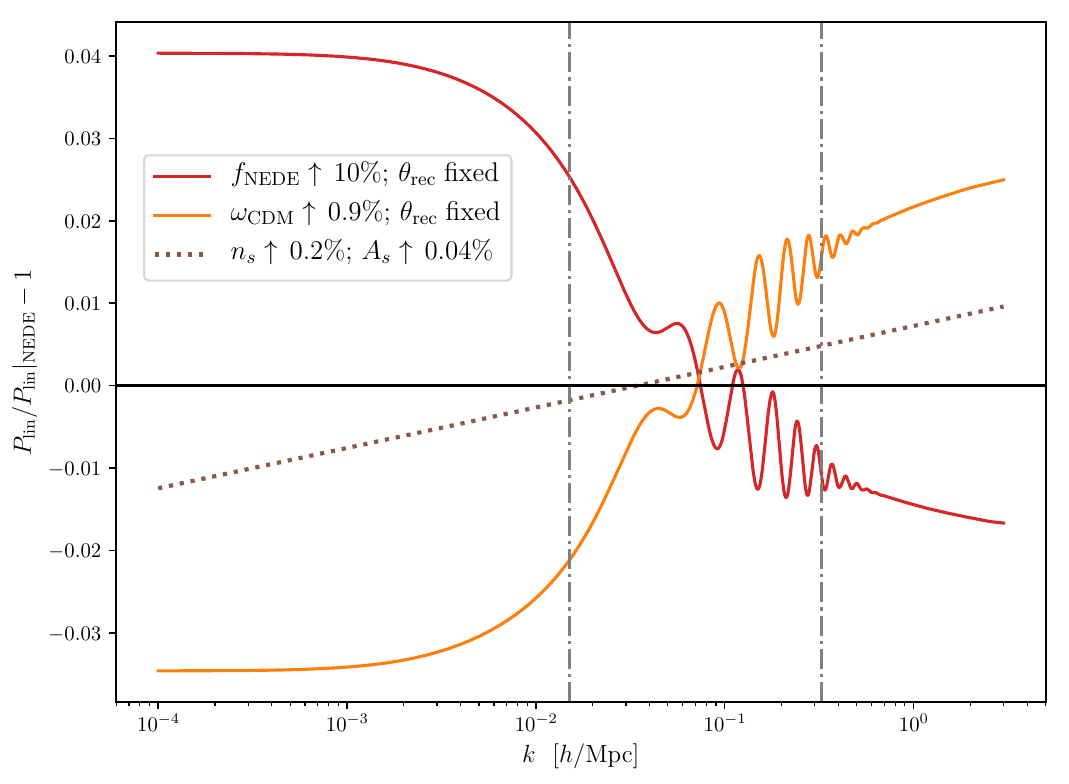}
         \caption{Effect of varying different NEDE parameters.}
         \label{fig:Pk_vary_params}
     \end{subfigure}
          \hfill
     \begin{subfigure}[b]{0.49\textwidth}
         \centering
         \includegraphics[width=\textwidth]{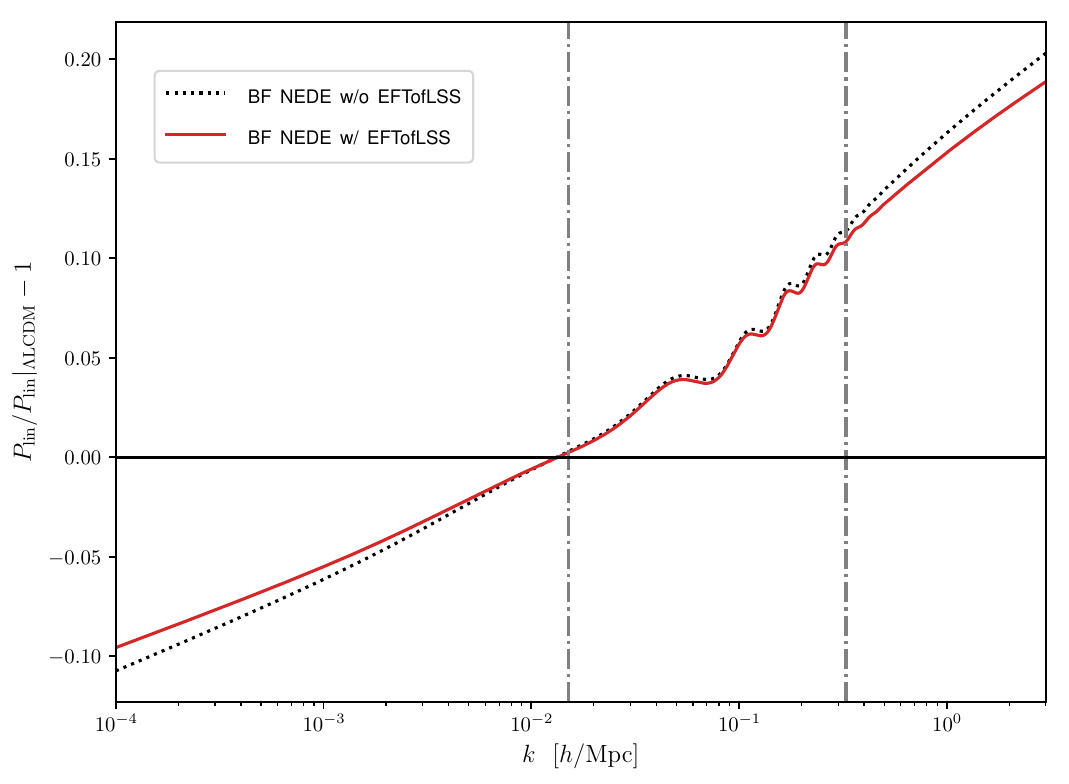}
         \caption{Comparison between bestfit $\Lambda$CDM and NEDE.}
         \label{fig:Pk_compare}
     \end{subfigure}
        \caption{Relative change of the linear matter power spectrum. Bestfit (BF) values are taken from Tab.~\ref{tab:means_NEDE_LCDM}. The $k$ range between the vertical dash-dotted lines makes a $\simeq 90\%$  contribution to the $\sigma_8$ integral~\eqref{eq:sigma8} (when evaluated for $P_\mathrm{lin}$). Generically, NEDE leads to less power on large and more power on small scales, causing a small net increase in $\sigma_8$. } %(in units $[\text{Mpc}/h]^3$)
        \label{fig:Pk_NEDE}
\end{figure}

The LSS tension is often quantified in terms of the $\sigma_8$ parameter, defined as the root-mean-square mass fluctuation within a sphere of radius $8  \, \textrm{Mpc}/h$, 
\begin{align}\label{eq:sigma8}
\sigma_8^2 = \frac{1}{2\pi^2 }\int dk k^2 P( k) \left[W(k \times 8  \, \textrm{Mpc}/h)\right]^2,
\end{align}
where $W(x) = (3/x) [\sin(x) /x^2 - \cos(x)/x]$ is the Fourier transformation of a top-hat window function and $P(k)$ is the matter power spectrum today. In short, the $\Lambda$CDM inferred value of $\sigma_8$ tends to be too large, leading to a $\simeq \, 2.5 \sigma $ tension when looking at measurements of say $S_8 = \sigma_8 \sqrt{\Omega_m / 0.3}$~(see for example \cite{Joudaki:2019pmv} and references in \cite{DiValentino:2020vvd}).

NEDE (as well as other early dark energy models) is known to slightly increase $\sigma_8$ and hence not reduce tensions with LSS data when simultaneously fitted to CMB data.  To explain this, we first need to understand how NEDE preserves the fit to CMB data (for more details see the discussion in \cite{Lin:2019qug} and \cite{Niedermann:2020dwg} in the case of acoustic early dark energy and NEDE, respectively). We already stated the main mechanism. Increasing $f_\textrm{NEDE}$ lowers the sound horizon $r_s(z_\text{rec})$. In order to keep the corresponding (highly constrained) angular scale $\theta_\mathrm{rec}$ fixed, $H_0$ needs to be increased.  Aside from this background effect, NEDE also manifests itself on the level of perturbations: the decaying NEDE fluid supports its own acoustic oscillations, which due to their positive pressure lead to a quicker decay of the gravitational potential that drives CMB oscillations. This potentially dangerous effect can then be reversed by increasing the cold dark matter density through $\omega_\textrm{cdm}$. Moreover, the background modification leads to a reduction of the CMB damping scale, which needs to be countered by increasing both the amplitude $A_s$ and tilt $n_s$ of the primordial power spectrum. While all these effects can be balanced in a way which leaves the CMB power spectrum approximately invariant, this is not exactly true for the matter power spectrum. 

This is illustrated in Fig.~\ref{fig:Pk_vary_params}, where we vary different model parameters to study their impact on the \textit{linear} matter power spectrum  (relative to the bestfit cosmology in the third column of Tab.~\ref{tab:means_NEDE_LCDM}). The relative changes are chosen such that the accumulated effect of all parameter changes on the CMB power spectrum is approximately vanishing.\footnote{This is illustrated explicitly in Fig.~9 in \cite{Niedermann:2020dwg} for the same parameter choices.} Moreover, in order to account for the main degeneracy between $f_\textrm{NEDE}$ and $H_0$ we kept $\theta_\mathrm{rec}$ fixed in all plots by dialing the value of $H_0$ correspondingly. To keep the discussion simple, we focus on scales that give the dominant contribution $(90\, \%)$ to the $\sigma_8$ integral in \eqref{eq:sigma8} and are demarcated by the dash-dotted lines.\footnote{On larger scales, the main effect is caused by changes to the background evolution.} First, the red line shows the effect of an $10\%$ increase in $f_\textrm{NEDE}$. This leads to a loss of power on small scales, which, again, can be attributed to the quicker decay of the gravitational potential. Now, the CMB fit can be preserved by increasing $\omega_\textrm{cdm}$ (by $\simeq 0.9\%$, yellow line), $A_s$ (by $\simeq 0.04\%$, dotted purple line) and $n_s$ (by $\simeq 0.2\%$, dotted purple line). However, we see that this cancelation does not work quite as perfectly in the case of the matter power spectrum. Instead, the additional power due to the increase in cold dark matter and primordial power over-compensates the depressing effect of $f_\textrm{NEDE}$, leaving us with slightly more power on  BAO scales (and less power on larger scales). This is also shown in the right panel when comparing our NEDE bestfit cosmology with $\Lambda$CDM: NEDE is changing the shape of the linear power spectrum by tilting it. As a side effect, there is more power on scales that were  sub-horizon by the time of matter-radiation equality, leading to a small increase in $\sigma_8$ through \eqref{eq:sigma8}. This discussion also shows that LSS data is going to be a challenge for NEDE.  We will therefore assess its constraining power in the next section by applying the EFTofLSS to BOSS/SDDS data. This analysis probes changes in the shape of the power spectrum on BAO scales and hence should be affected by the observed tilting.

\section{Data analysis and results}

\begin{figure*}
 \centering 
{\includegraphics[width=14.5cm]{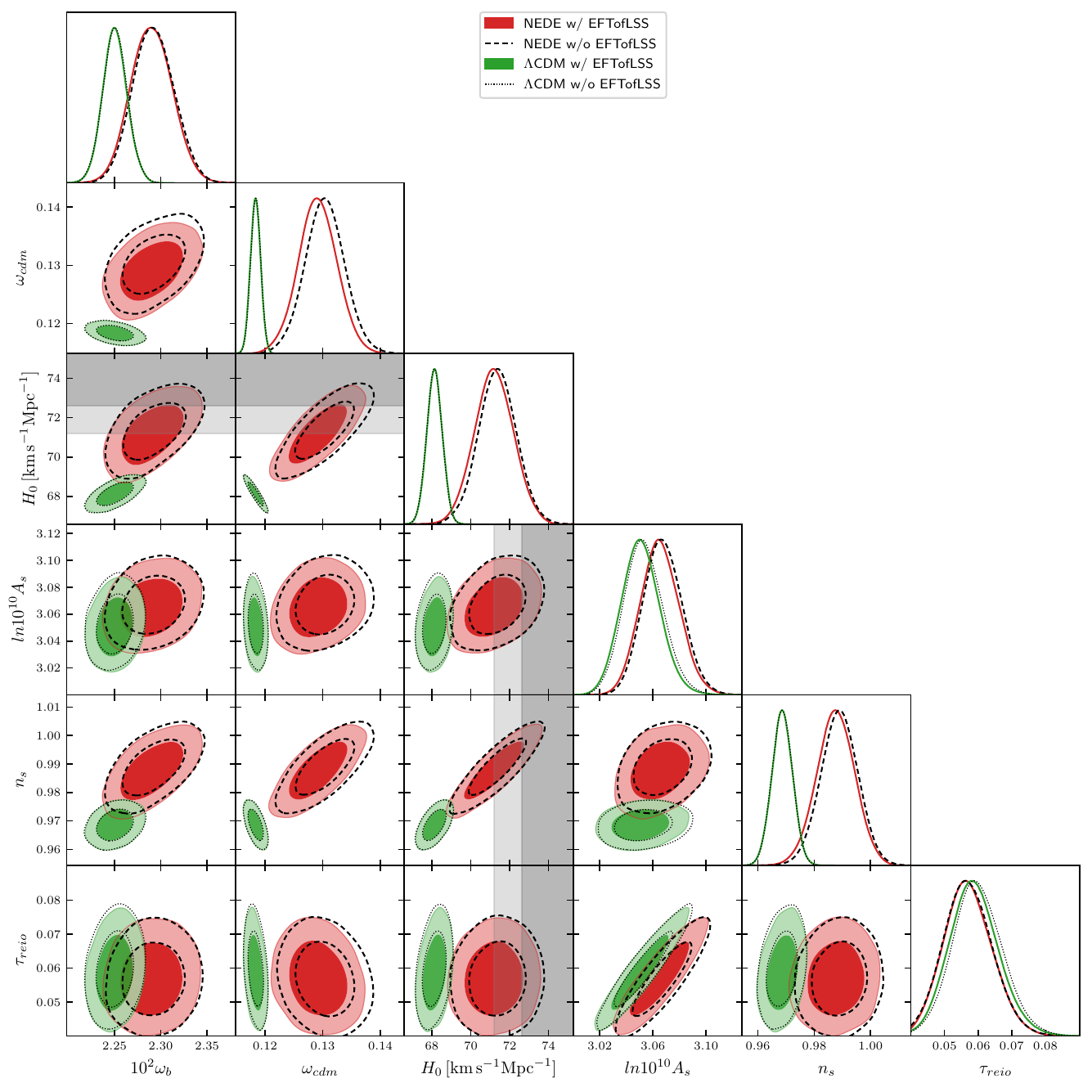}}%
 \caption{Posteriors and covariances of standard cosmological parameters for $\Lambda$CDM (green) and NEDE (red). The result of the combined analysis without  EFTofLSS corresponds to the dashed (NEDE) and dotted ($\Lambda$CDM) contours. Here and henceforth, the $68 \%$ C.L.\ and the $95 \%$ C.L. correspond to the darker and lighter shaded regions, respectively. The gray band corresponds to the SH$_0$ES constraint on $H_0$. Overall, including EFTofLSS has a negligible effect. }
\label{fig:triangle}
\end{figure*}

We use the publicly available code {\tt TriggerCLASS}\footnote{\url{https://github.com/flo1984/TriggerCLASS}} \cite{Niedermann:2020dwg}, which implements NEDE in the Boltzmann code {\tt CLASS} (Cosmic Linear Anisotropic Solving System)~\cite{Blas:2011rf}. We then scan the cosmological parameters with the Monte Carlo Markov Chain (MCMC) code {\tt MontePython}~\cite{Audren:2012wb,Brinckmann:2018cvx}, employing the Metropolis-Hastings algorithm. 
To that end, we impose flat priors with standard ranges on the dimensionless baryon and cold dark matter density, $\omega_b$ and $\omega_{cdm}$, the Hubble parameter $H_0$, the amplitude of primordial curvature perturbations at $k=0.05\,\text{Mpc}^{-1}$, $\ln 10^{10} A_s$, the spectral tilt $n_s$, and the reionization optical depth $\tau_\text{reio}$. In keeping with the {Planck} convention, the neutrino sector contains two massless and one massive species with $M_\nu = 0.06 \, \text{eV}$, where the effective number of relativistic degrees of freedom is fixed to $N_\text{eff} = 3.046$. Alongside these standard parameters, we also vary the trigger mass through $\log_{10}(m/m_0)$ (where $m_0 = 1/\text{Mpc}$) and the NEDE fraction $f_\text{NEDE}$. We impose flat priors with ranges $1.3 < \log_{10}(m/m_0) < 3.3$  and  $0<f_\text{NEDE}<0.3$. $\Lambda$CDM is recovered for $f_\text{NEDE} = 0$. For our base model we further fix $H(t_*) / m =0.2$ in accordance with our theoretical discussion. We also set $w_\text{NEDE}(t_*) =2/3 $ corresponding to an initial admixture of a stiff fluid.\footnote{For extended runs where either parameter is allowed to vary see~\cite{Niedermann:2020dwg}.} The perturbation sector assumes a rest-frame sound speed that equals the adiabatic sound speed and a vanishing viscosity parameter (as defined in \cite{Hu_1998}).  To summarize, we study a two-parameter extension of $\Lambda$CDM with flat parameter priors.
 
In all our runs, we include the following data sets: The {Planck 2018} TT, TE, EE and lensing likelihood with the full set of nuisance parameters~\cite{Aghanim:2019ame}, supernovae data from the combined {Pantheon} sample~\cite{Scolnic:2017caz}, the locally measured value $H_0 = 74.03  \pm 1.42 \, \kmsMpc$ ($68 \%$ C.L.) from SH$_0$ES~\cite{Riess:2019cxk} implemented as a Gaussian prior, the primordial Helium abundance $Y_p = 0.2449 \pm 0.0040$ ($68 \%$ C.L.) from~\cite{Aver:2015iza} and the small-$z$ BAO measurements of the  {SDDS DR7} main Galaxy sample~\cite{MGS} and the 6dF Galaxy Survey~\cite{6dF} at redshifts $z = 0.15$ and $z=0.106$, respectively. With regard to any additional data set, we distinguish two different analyses:

\begin{enumerate}
\item Combined analysis \textbf{without EFTofLSS}: High-$z$ BAO measurements together with constraints on $f \, \sigma_8$, quantifying the growth of structure, obtained from the {CMASS} and {LOWZ} galaxy samples of {BOSS DR 12}~\cite{Alam:2016hwk} at redshifts $z= 0.38$, $0.51$, $0.61$. This run does not capture the full shape of the matter power spectrum (it merely contains some condensed information  through $f \sigma_8$). 

\item Combined analysis \textbf{with EFTofLSS}: Here, we include the EFTofLSS applied to the {BOSS/SDDS} sample~\cite{DAmico:2019fhj,Ivanov:2019pdj,Colas:2019ret}. It contains the full-shape information on the galaxy power spectra obtained from the sky-cuts CMASS NGC, CMASS SGC and LOWZ NGC at the effective redshifts $z= 0.57$ (CMASS) and $z=0.32$ (LOWZ). This is combined with constraints on BAO parameters measured from the same samples using the post-reconstructed power spectra and taking into account the covariance among all data sets. The full-shape and BAO data together with their covariances are implemented via the {\tt MontePython} likelihood extension {\tt PyBird}\footnote{\url{https://github.com/pierrexyz/pybird.git}}. {We impose the same priors on the eight EFT parameters as the ones in \cite{DAmico:2020kxu}, including a flat prior on the linear galaxy bias $b_1$.}
\end{enumerate}

 \begin{table}
\centering
\begin{footnotesize}
\begin{tabular}{|c|c|c|c|c|} 
 \hline \hline
{\bf Parameter} 			&  \multicolumn{2}{c|}{$\Lambda$CDM} 												&  \multicolumn{2}{c|}{NEDE}  \\ 
 					& w/o EFTofLSS 							& w/ EFTofLSS 	 						&  w/o EFTofLSS 						&  w/ EFTofLSS 	 \\ 
\hline \hline
$100~\omega_{b }$ 		&$2.251_{-0.013}^{+0.014}$ ($2.251$)			&$2.250_{-0.014}^{+0.013}$ 	($2.256$)		& $2.292_{-0.024}^{+0.022}$  ($2.297$) 		&  $2.290_{-0.023}^{+0.022}$ ($2.288$)\\ 
$\omega_{cdm }$ 		&$0.1184_{-0.0009}^{+0.0009}$ ($0.1183$)		&$0.1183_{-0.0009}^{+0.0009}$ ($0.1181$)	& $0.1304_{-0.0035}^{+0.0034}$  ($0.1306$)	& $0.1291_{-0.0034}^{+0.0033}$ ($0.1295$)\\
$H_0 \, [\kmsMpc]$ 		&$68.13_{-0.41}^{+0.41}$ ($68.16$)				&$68.14_{-0.41}^{+0.40}$ ($68.31$)			& $71.4_{-1.0}^{+1.0}$ ($71.5$)			& $71.2_{-1.0}^{+1.0}$ ($71.28$)\\
$ln10^{10}A_{s }$ 		&$3.053_{-0.016}^{+0.014}$ ($3.053$)			&$3.051_{-0.015}^{+0.014}$ ($3.049$)		& $3.067_{-0.015}^{+0.014}$ ($3.068$)		& $3.065_{-0.015}^{+0.014}$($3.064$)\\
$n_{s }$ 				& $0.9686_{-0.0037}^{+0.0037}$ ($0.9698$)		& $0.9686_{-0.0037}^{+0.0037}$ ($0.9696$) 	& $0.9889_{-0.0066}^{+0.0067}$  ($0.9912$)	&  $0.9876_{-0.0066}^{+0.0070}$($0.9884$)\\ 
$\tau_{reio }$ 			&$0.0599_{-0.0078}^{+0.0071}$ ($0.0598$)		&$0.0589_{-0.0078}^{+0.0068}$ ($0.0573$)	& $0.0571_{-0.0077}^{+0.0068}$  ($0.0572$)	&  $0.0571_{-0.0077}^{+0.0068}$ ($0.0557$)\\ 
$f_\text{NEDE }$ 		&--										&--									& $0.126_{-0.029}^{+0.032}$  ($0.1296$)		& $0.117_{-0.030}^{+0.033}$ ($0.120$)\\ 
$\log_{10}(m/m_0)$		&--										&--									& $2.56_{-0.10}^{+0.12}$ ($2.57$)			& $2.55_{-0.11}^{+0.12}$ ($2.53$)\\
\hline
$\sigma_8$ 			&$0.8090_{-0.0065}^{+0.0060}$ ($0.8092$)		& $0.8080_{-0.0063}^{+0.0058}$  ($0.8064$)	& $0.839_{-0.010}^{+0.010}$ ($0.841$)		& $0.836_{-0.010}^{+0.010}$ ($0.837$)\\ 
$S_8$ 				&$0.814_{-0.010}^{+0.010}$ 					& 	$0.813^{+0.010}_{-0.010}$			& {$0.841^{+0.012}_{-0.012}$} 		 		&  $0.836^{+0.012}_{-0.012}$\\ 
$r_s^{d} $ [Mpc] 		& $147.40_{-0.23}^{+0.23}$ ($147.38$)			& $147.40_{-0.23}^{+0.23}$	($147.39$)	&  $141.0_{-1.7}^{+1.6}$   ($140.9$)			& $141.6_{-1.7}^{+1.6}$ ($141.4$)\\ 
$z_{*} $				&--										&-- 									& $4920_{-730}^{+620}$ ($4960$)			&  $4900_{-800}^{+660}$ ($4720$)\\ 
\hline 
%Tension $S_8 	$		& $1.9 \, \sigma	$							&$1.9 \, \sigma	$						& $2.8 \, \sigma		$					& $2.7 \, \sigma		$ \\
$\Delta \chi^2 	$		& 0										&0									& -15.6								& -14.7\\
$f_{\text{NEDE}} \neq 0$	& --										& --									& $4.3 \, \sigma$						& $3.9 \, \sigma$\\ 
$10^3 \times \max{(R-1)} $&		2.1								&			2.7						&		7.8							& 3.3 \\
\hline \hline
 \end{tabular} \\ 
 \end{footnotesize}
\caption{ The mean value and $\pm 1 \,\sigma$ error (with bestfit value in parentheses) of the cosmological parameters from our combined analyses for $\Lambda$CDM and NEDE with and without EFTofLSS. }
\label{tab:means_NEDE_LCDM}
\end{table}

For each data set combination we run both the $\Lambda$CDM and our NEDE base model using between 8 and 16 chains. We consider chains to be converged if the Gelman-Rubin criterion~\cite{Gelman:1992zz} fulfills $R-1 < 0.01$. Especially in the case of NEDE, this requires a rather large number of total steps of the order $3 \times 10^6$. As we find, this is vital to deriving well-converged uncertainties for $f_\text{NEDE}$, needed to make a reliable statement about the statistical evidence for NEDE.   The exact convergence values are detailed in Tab.~\ref{tab:means_NEDE_LCDM}. As initial covariance matrices we use the ones from the respective $\Lambda$CDM runs, which are then updated through {\tt MontePython}'s `superupdate' option. We also tested our data pipeline, including its {\tt PyBird} implementation, by reproducing different results obtained in~\cite{DAmico:2020kxu} in the case of $w$CDM.\footnote{Specifically, we checked that we agree with the results of the MCMC analyses performed with the data set combinations  `BAO + FS', 'BAO + FS w/o Ly-$\alpha$', `CMB + BAO' and 'CMB + FS + BAO'. }

\begin{figure}
 \centering 
{\includegraphics[width=16.5cm]{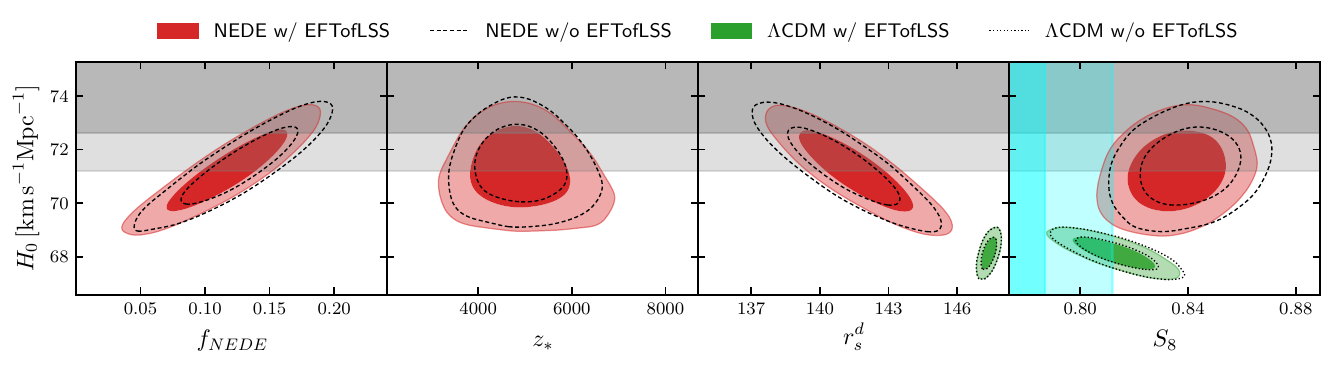}}%
 \caption{Covariances of $H_0$ vs a subset of parameters for the combined analysis with (red and green filled contour) and without (dashed and dotted contour) EFTofLSS.
 A recent constraint on $S_8$ from weak gravitational lensing~\cite{Joudaki:2019pmv} is depicted by the light blue band. The gray band represents the recent SH$_0$ES measurement. The change in $f_\text{NEDE}$ when including EFTofLSS is insignificant.}
\label{fig:NEDE_LCDM_LSS_rect}
\end{figure}

A detailed discussion of NEDE and its phenomenology has been provided in~\cite{Niedermann:2020dwg}. Here, we will therefore limit ourselves to a quick review of the main phenomenological features of NEDE and rather focus on the impact additional LSS data has on the extracted parameter values.  As mentioned before, the primary effect of NEDE is to lower the sound horizon, which is balanced by increasing $H_0$. This can be seen in Fig.~\ref{fig:NEDE_LCDM_LSS_rect}, which shows an (approximate) degeneracy between $H_0$ and the sound horizon at radiation drag $r_s^d$. As a result, the NEDE contour (w/ and w/o EFTofLSS) largely overlaps with the gray band representing the SH$_0$ES measurement, thereby resolving the Hubble tension. Another crucial effect of NEDE, especially relevant for LSS, is its positive correlation with the (dimensionless) dark matter energy density $\omega_\text{cdm}$, as explained through Fig.~\ref{fig:Pk_vary_params}. This can, for example, be seen in the $H_0$ vs $w_\text{cdm}$ plot in Fig.~\ref{fig:triangle}. Finally, the enhanced diffusion damping on small scales is counter-acted by a reduced spectral tilt (or $n_s  \to 1$ equivalently). {Specifically}, from Fig.~\ref{fig:triangle} we infer that $n_s$ becomes $2 \, \sigma$ compatible with a scale invariant spectrum. 

Here, we ask whether this picture is adversely affected by including additional LSS data as recently claimed in the context of the old EDE proposal~\cite{Ivanov:2020ril, DAmico:2020ods}. The short answer is that this is not the case. Including additional LSS data only leads to an insignificant change of previous results without the EFTofLSS data set. This is obvious from comparing the red (w/ EFTofLSS) and dashed contours (w/o EFTofLSS) in Fig.~\ref{fig:triangle} and \ref{fig:NEDE_LCDM_LSS_rect}, which are almost identical. On a more quantitative level, we report a small reduction of the preferred amount of NEDE as detailed in Tab.~\ref{tab:means_NEDE_LCDM}. Without including the EFTofLSS, we have $f_\text{NEDE} = 12.6^{+3.2}_{-2.9}\, \%$ which corresponds to a $4.3\, \sigma$ evidence for a non-vanishing NEDE parameter and an increased Hubble parameter of $H_0 = 71.4\pm 1.0\, \kmsMpc$.   These values undergo small insignificant ($<0.5 \, \sigma$) changes  when we include the EFTofLSS, specifically  $f_\text{NEDE} = 11.7^{+3.3}_{-3.0}\, \%$ and $H_0 = 71.2 \pm 1.0 \, \kmsMpc$, corresponding to a (still large) $3.9\, \sigma$ evidence for NEDE. With regard to the standard $\Lambda$CDM parameters, the biggest change occurs for $w_\text{cdm}$, which decreases by $1 \%$ whereas the other $\Lambda$CDM parameters change by less than $0.3 \%$. This had to be expected as $\omega_\text{cdm}$ directly affects LSS parameters and hence is most constrained by the EFTofLSS. Moreover, {as argued in Sec.~\ref{sec:sigma8}}, decreasing $w_\text{cdm}$, while keeping the other parameters approximately constant, leads to a slight decrease in the $\sigma_8$ parameter. This in turn reduces $S_8 = \sigma_8 \sqrt{\Omega_m / 0.3}$ by $0.6 \, \%$ (the effect is attenuated by a slight drop in $H_0$). The phenomenological constraints are depicted as the light blue band in the last panel of Fig.~\ref{fig:NEDE_LCDM_LSS_rect} (we cite $S_8 = 0.762^{+0.025}_{-0.024}$ from a combined tomographic weak gravitational lensing analysis of the Kilo Degree Survey and the Dark Energy Survey~\cite{Joudaki:2019pmv}). We then find that the $S_8$ tension is still significant at $ 2.7 \, \sigma$ (and $2.8 \,\sigma$ without EFTofLSS). However, this has to be compared to a $\simeq \, 2.5 \sigma$ tension within $\Lambda$CDM~\cite{Joudaki:2019pmv, Niedermann:2020dwg}  (without SH$_0$ES)\footnote{Note that including SH$_0$ES lowers the tension within $\Lambda$CDM. This however is not a viable way of alleviating the problem as it relies on combining incompatible data sets.}, which is only marginally lower. 

\begin{table}
\centering
\begin{footnotesize}
\begin{tabular}{|l|c|c|c|c|c|c|c|} 
 \hline 
 \hline
{\bf Dataset} & \multicolumn{2}{c|}{$\Lambda$CDM} & \multicolumn{4}{c|}{NEDE} \\ 
\hline
\multicolumn{1}{|r|}{EFTofLSS + BAO} 		& w/o 			&w/		&\multicolumn{2}{c|}{w/o}		& \multicolumn{2}{c|}{w/} \\  
\hline
\multicolumn{1}{|r|}{} 				&$\chi^2$ 		& $\chi^2$ 	& $\chi^2$& $\Delta \chi^2$ 	&  $\chi^2$ 	& $\Delta \chi^2$ 	 \\  
							 \hline\hline
{\emph{Planck}} high-$\ell$ TT, TE, EE & 2,348.9 	&  2351.1		& 2,348.2 	&-0.8			&	2348.2	&-2.9	\\  
{\emph{Planck}} low-$\ell$ TT		& 22.7 		&  22.6		& 20.8 	&-1.9			&	21.0		&-1.6	\\  
{\emph{Planck}} low-$\ell$ EE 		& 397.3 		& 396.5		& 396.4 	&-0.9			&	396.1	&-0.4	\\  
{\emph{Planck}} lensing 			& 9.1 		&  9.6		& 9.5 	&0.4				&	9.4		&-0.2	\\  
 BAO low-z					& 1.6 		&  1.8		& 1.8 	&0.1				&	1.8		& 0.1	\\  
 BAO high-z  + $f \sigma_8$		& 5.9 		&  --			& 6.8 	&0.9				&	--		& --	\\  
 EFTofLSS + BAO / CMASS NGC	& --			& 65.9		& --	 	&--				&	67.1 		& 1.3	\\  
 EFTofLSS + BAO / CMASS SGC	& --			& 61.9		& --	 	&--				&	63.0		& 1.1	\\  
 EFTofLSS + BAO / LOWZ NGC	& --			& 69.9		& --	 	&--				&	69.9 		& 0.0	\\  
 Pantheon 					& 1,027.0 		& 1026.9		&1,027.3 	&0.3				&	1027.4 	&0.5	\\  
 BBN 						& $<0.1$ 		& $<0.1$ 		&$<0.1$ 	&0.0				&	$<0.1$	&0.0	\\  
 SH$_0$ES 					& 17.1 		& 16.2		& 3.3		&-13.8			&	3.7		& -12.5	\\ \hline \hline 
 $\chi^2$(total) 					& 3,829.6 		&  4,022.5		& 3,814.0 	&--				& 	4,007.8	& --	\\
 $\Delta \chi^2 $(total)			& -- 			& --			&-- 		&-15.6			& --			&-14.7	\\ \hline \hline
\end{tabular} \\
\end{footnotesize}
\caption{The bestfit $\chi^2 = - 2 \ln(\mathcal{L})$ from combined analysis with and without EFTofLSS. The relative fit improvement is quantified through $\Delta \chi^2 = \chi^2 (\text{NEDE}) - \chi^2(\Lambda\text{CDM})$.   }
\label{tab:chi2}
\end{table}

This picture is confirmed by our $\chi^2$ analysis in Tab.~\ref{tab:chi2}. It shows that the overall $\chi^2$ improvement is only slightly affected by including the EFTofLSS and amounts to $\Delta \chi^2 (\text{total}) \simeq -15$.  In this context, it is interesting to note that the $\Lambda$CDM fit to high-$\ell$ Planck data gets worse upon inclusion of the EFTofLSS data, which we believe is a manifestation of the LSS/$S_8$ tension already present within $\Lambda$CDM at the $\simeq 2-3 \, \sigma$ level~\cite{Joudaki:2019pmv,Asgari:2019fkq}. Within NEDE, on the other hand, that deterioration of the high-$\ell$ Planck fit can be completely avoided. We attribute this to the tendency of NEDE to give less power on small scales, which is a distinctive feature when comparing with its EDE competitors and $\Lambda$CDM (see Fig.~17 in \cite{Niedermann:2020dwg}). This, however, comes at the price of worsening the fit to the EFTofLSS data set. Both effects -- the improved fit to Planck and the worse fit to the full-shape data -- compensate each other almost perfectly. In other words, the additional LSS data leads to a similar overall (negative) effect on both the $\Lambda$CDM and NEDE fit, which explains why it cannot significantly lower the evidence for NEDE.

\begin{figure}
 \centering 
{\includegraphics[width=16.5cm]{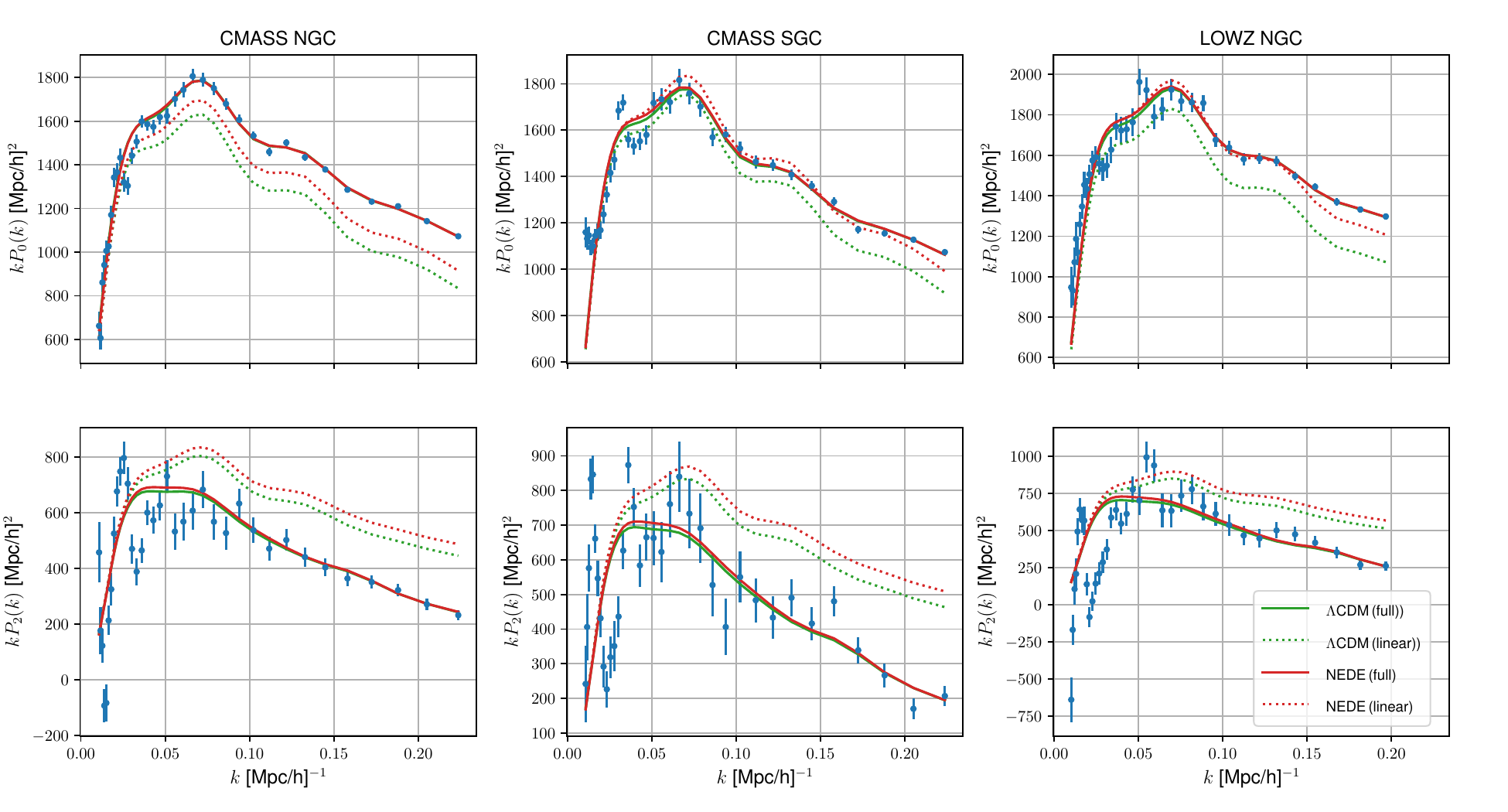}}%
 \caption{{Monopole (first row) and quadrupole (second row) of the EFT-corrected redshift-space power spectrum for all three skycuts. The  green and red curves correspond respectively to the $\Lambda$CDM and NEDE bestfit cosmology with the EFTofLSS as detailed in Tab.~\ref{tab:means_NEDE_LCDM}. The dotted line is the contribution from the linear power spectrum. In agreement with our $\chi^2$ comparison in Tab.~\ref{tab:chi2}, $\Lambda$CDM and NEDE provide similiarly good fits.}}
\label{fig:bestfit}
\end{figure}

{Finally, we plotted the bestfit monopole ($\ell =0$) and quadrupole  ($\ell =2$)  of the EFT-corrected redshift space power spectrum in Fig.~\ref{fig:bestfit}. More explicitly, they are obtained from}
\begin{align}
P^\mathrm{(true)}_\ell(q) = \frac{2 \ell +1}{2} \int_{-1}^1 \mathrm{d}\mu P(q, \mu) \mathcal{L}_\ell(\mu)\,,
\end{align}
after applying the Alcock-Paczynski transformation to account for the fact that the observation uses a fictitious cosmology to convert redshifts and celestial coordinates to Cartesian coordinates. Here, $\mu = \mathbf{k} \cdot \hat{\mathbf{z}}/k$ with $\hat{\mathbf{z}}$ the line-of-sight unit vector, $\mathcal{L}_\ell$ are the Legendre polynomials, and $P(q, \mu)$ is the redshift-space power spectrum at one-loop order. Its expression in terms of EFT parameters can be found for example in the appendix of \cite{DAmico:2020kxu} alongside the explicit Alcock-Paczynski transformation relating $\{q,\, P_\ell^\mathrm{(true)}(q)\}$ with $\{k, P_\ell(k)\}$. To provide an explicit example, the linear contribution to $P(q, \mu)$, giving rise to the dotted lines in Fig.~\ref{fig:bestfit}, is given by 
\begin{align}
 P_\mathrm{lin}(q, \mu) = \left( b_1 + f \mu^2\right)^2 P_\mathrm{lin}(q)\;,
\end{align}
{where $f$ is the linear growth rate, $P_\mathrm{lin}(q)$ the linear power spectrum at a given redshift, and $b_1$ the linear galaxy bias, which we fitted for each skycut separately. Again, the crucial observation is that both NEDE and $\Lambda$CDM fit the full shape of the power spectrum equally well. Moreover, in agreement with Fig.~\ref{fig:Pk_compare}, we see that the linear spectrum predicts more power on short scales for NEDE due to the relative tilting we observed before. This difference, however, is absent in the loop-corrected expressions. In other words, the EFT corrections act in favor of NEDE. This discussion also shows that the scatter and error in the data is still too large to discriminate between the tiny shape difference between the $\Lambda$CDM and NEDE spectrum. }

\section{Conclusions}

In this short note, we confronted NEDE with full-shape LSS data using the EFTofLSS applied to BOSS/SDSS in order to address recent concerns regarding the phenomenological viability of a class of EDE models. To that end, we used the publicly available code {\tt PyBird}, which allowed us to implement the same data pipeline as the one used in \cite{DAmico:2019fhj} to constrain single-field EDE models.\footnote{This work does not include direct constraints on $S_8$ from photometric surveys to retain comparability with the {\tt PyBird} analysis of EDE in~\cite{DAmico:2019fhj}. We intend to study the effect of further LSS constraints in our future work when looking at extensions of NEDE that can alleviate the LSS tension. In contrast to the EFTofLSS analysis, which takes the linear power spectrum as input, this requires explicit N-body simulations to test the accuracy of  semi-analytic methods like \texttt{Halofit}~\cite{Smith:2002dz} or \texttt{HMcode}~\cite{Mead:2015yca}.}    We report that adding the full-shape information has an insignificant effect on NEDE. In particular, we still find a rather high $\simeq 4 \, \sigma$ evidence for a non-vanishing fraction of NEDE alongside $H_0 = 71.2 \pm 1.0 \, \kmsMpc$ ($68 \%$ C.L.), which is fully compatible with $H_0 = 71.4\pm 1.0\, \kmsMpc$ ($68 \%$ C.L.) obtained without full-shape data. In conclusion our model is consistent with current LSS data as implemented by {\tt PyBird}.

However, it is also clear from our analysis that NEDE cannot improve on the tension with LSS data already present in $\Lambda$CDM. In particular, NEDE is $2.7\, \sigma$ discrepant with the value of $S_8$, which is of a similar level as the tension within $\Lambda$CDM. There are two main conclusions we can reach from this. First, more LSS data, as it is provided by present and future spectroscopic galaxy surveys such as Euclid~\cite{Amendola:2016saw} and the Dark Energy Spectroscopic Instrument~\cite{Levi:2019ggs},  has the potential to confirm or rule out the base NEDE model as a resolution to the Hubble tension.  A similar point was made in the case of single-field EDE~\cite{Ivanov:2020ril}. Second, further improvements of the current model should be guided by the aim to reduce the $S_8$ tension below its $\Lambda$CDM level. For different ideas how this can be achieved within NEDE see the corresponding discussion in~\cite{Niedermann:2020dwg}.

Finally, we have limited the discussion here to the base NEDE model for simplicity. A more extensive LSS analysis which allows the fluid parameters to vary freely might provide additional freedom needed to better accommodate LSS data.

\acknowledgments
 
We thank Guido D'Amico for private correspondence, assistance in setting up the base $\Lambda$CDM run using {\tt PyBird} and comments on the draft. {We also thank the SDU eScience Center for providing us with the computational resources for the MCMC analysis.}
This work is supported by Villum Fonden grant 13384. 

\bibliography{NewEDE_and_LSS}

\end{document}